# Comparison of Lubrication Approximation and Navier-Stokes Solutions for Dam-Break Flows in Thin Films

Sandy Morris, Mathieu Sellier[1], Abdul Rahman Al Behadili

Mechanical Engineering Department, University of Canterbury, Christchurch 8140, New Zealand

**ABSTRACT**

*The levelling of a thin liquid film has been analysed experimentally, theoretically, and numerically. The purpose of this contribution is to compare solutions from two different finite element models: one based on the 2D Navier-Stokes equations, and the other based on a simplification thereof, known as the lubrication approximation. The latter is significantly less computationally expensive, hence it is desirable to know in which parameter range this approximation is valid. As a benchmark problem, we consider the classical dam-break problem. This paper establishes the validity window of the lubrication approximation in terms of the Reynolds number, Froude number, and film aspect ratio. To validate the solutions produced by the solver, experimental data has been obtained for the levelling of silicone oil. The free surface velocity measured experimentally is compared to that produced by the two numerical models and found to have relatively good agreement. The lubrication approximation is then compared with the Navier-Stokes model for a range of Reynolds and Froude numbers, and for a range of film aspect ratios. The results show that the lubrication approximation is remarkably robust and particularly accurate for low Reynolds and Froude numbers, for which inertial effects are negligible. Using fixed Reynolds and Froude numbers of 100 and 0.01 respectively, the lubrication approximation is accurate for aspect ratios up to 1%.*

**INTRODUCTION**

Measuring the rheology of a fluid can prove a challenging task. The fluid may, for example, be in too small a quantity or simply be too hot for standard rheometer. It has recently been proposed that the rheology of a fluid can be inferred indirectly by measuring the free surface response of the flow and comparing it to model predictions [1-4]. Dam-break experiments have been performed whereby a quantity of fluid held by a gate is suddenly released and let to spread freely [5-8]. Such flows can be modelled using the full Navier-

---

[1] To whom correspondence should be addressed: mathieu.sellier@canterbury.ac.nz

Stokes equations or a well-known approximation known as the lubrication approximation. The lubrication approximation leads to a set of equations which are simpler to solve and less computationally expensive than the full Navier-Stokes equations. It is however poorly quantified when this approximation holds true. The key objective of this paper is to compare solutions of the lubrication approximation with those from the full Navier-Stokes equations and with experimental results, and establish the validity of this approximation.

The lubrication approximation or long-wave approximation has been commonly applied to studies on the dynamics of thin liquid films. Excellent reviews on the evolution of thin films can be found in [9,10].

Few studies have been carried out on the validity of the lubrication approximation in comparison to the full Navier-Stokes solution, despite its wide application. One study that compares the two models in detail is [11], which analyzed the steady-state flow of a liquid over various stepped topographies, and quantified the error of the lubrication approximation over a range of Reynolds numbers and step heights. Reference [12] compares the lubrication approximation and Navier-Stokes solutions for the spreading of droplets. The authors found good quantitative agreement between the two approaches but the authors limited their study to low Reynolds numbers in order to isolate the error induced by the small slope assumption in the lubrication approximation.

This paper solves the dam-break problem, where a fixed volume of fluid is separated and kept at two different levels by a gate which is suddenly removed. The dam-break problem has been studied extensively in the past [5, 13]. Finite volume techniques have been used to solve the full Navier-Stokes equations [14]. Here, we use a finite element technique to solve the Navier-Stokes equations, with a similar method to that used in [15]. The lubrication approximation has previously been applied to the dam-break problem specifically [16,17], where the results from the model are compared with experimental data and found to have good agreement for the specific flow conditions analyzed. The problem of the conditions at which this agreement holds, however, is not addressed. Reference [13] compares the lubrication approximation for the dam-break problem with experimental data and similarity solutions for the Navier-Stokes equations. The focus is to compare the results at different flume inclines; the result is that different scalings in the lubrication approximation are required for horizontal, shallow slopes and large slopes. Reference [18] investigates the influence of the power-law rheological parameters on the free levelling dynamics during the dam-break in the context of the lubrication approximation.

The focus of this paper is to compare the lubrication approximation with solutions from the

Navier-Stokes solver for the horizontal dam-break problem, in order to establish the validity of the approximation in terms of dimensionless flow parameters and film aspect ratio. Experimental data has also been used to confirm the validity of both models.

Section 2 details the governing equations and the non-dimensionalization of the problem. Section 3 summarizes the finite element method of the numerical simulations used to solve the governing equations. The numerical simulations are validated against experimental data in Section 4. Section 5 presents the results and discussion of comparing the lubrication approximation with full 2D Navier-Stokes solutions. Conclusions are drawn in Section 6.

## MATHEMATICAL MODELS

Figure 1 shows a diagram of the dam-break problem before the gate is removed at $t = 0$. The height and length scales of the problem are defined as $H_0 = \frac{H_1 + H_g}{2}$ and $L_0 = L_1 + L_g$ respectively. The overall aspect ratio of the film is defined as $\epsilon = H_0/L_0$.

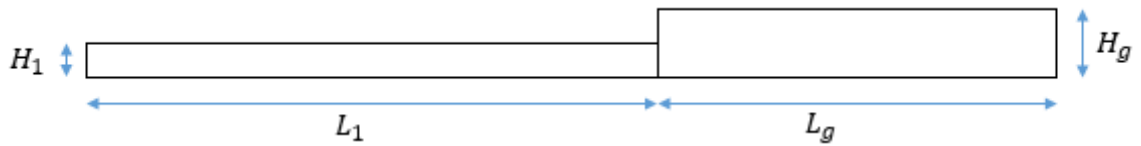

*Figure 1. Schematic of dam-break problem*

### Lubrication Approximation Governing Equation Derivation

The 2D Navier-Stokes equations for an incompressible Newtonian fluid can be defined as:

$$\rho \left( \frac{\partial u}{\partial t} + u \frac{\partial u}{\partial x} + w \frac{\partial u}{\partial z} \right) = -\frac{\partial p}{\partial x} + \mu \left( \frac{\partial^2 u}{\partial x^2} + \frac{\partial^2 u}{\partial z^2} \right), \tag{1}$$

$$\rho \left( \frac{\partial w}{\partial t} + u \frac{\partial w}{\partial x} + w \frac{\partial w}{\partial z} \right) = -\frac{\partial p}{\partial z} + \mu \left( \frac{\partial^2 w}{\partial x^2} + \frac{\partial^2 w}{\partial z^2} \right) - \rho g, \tag{2}$$

$$\frac{\partial u}{\partial x} + \frac{\partial w}{\partial z} = 0. \tag{3}$$

For low Reynolds number and unidirectional flow, to leading order in $\epsilon$, (1) and (2) reduce to:

$$\frac{\partial p}{\partial x} = \mu \frac{\partial^2 u}{\partial z^2}, \tag{4}$$

$$\frac{\partial p}{\partial z} = -\rho g. \tag{5}$$

Boundary conditions of the problem are given by the no-slip condition at the substrate, along

with no-shear and the pressure condition at the free surface:

$$u = w = 0 \quad \text{on } z = 0, \quad (6)$$

$$\frac{\partial u}{\partial z} = 0 \quad \text{on } z = h, \quad (7)$$

$$p = -\sigma \frac{\partial^2 h}{\partial x^2} \quad \text{on } z = h. \quad (8)$$

The kinematic boundary condition at the free surface is given by eq. (9). Dynamic boundary conditions are implemented using the balance of tangential and normal stresses at the free surface, by (10) and (11) respectively.

$$w = u \frac{\partial h}{\partial x} \quad \text{on } z = h \quad (9)$$

$$2\frac{\partial h}{\partial x}\left(\frac{\partial w}{\partial z} - \frac{\partial u}{\partial x}\right) + \left(1 + \left(\frac{\partial h}{\partial x}\right)^2\right)\left(\frac{\partial u}{\partial z} + \frac{\partial w}{\partial x}\right) = 0 \quad \text{on } z = h \quad (10)$$

$$\mu\left(\frac{\partial h}{\partial x}\left(\frac{\partial u}{\partial z} + \frac{\partial w}{\partial x}\right) - 2\frac{\partial w}{\partial z}\right) + p - p_0 + \sigma \frac{\frac{\partial h}{\partial x}}{(1 + \left(\frac{\partial h}{\partial x}\right)^2)^{3/2}} = 0 \quad \text{on } z = h \quad (11)$$

Integrating (4) twice with respect to z, and applying boundary conditions (6) and (7) gives:

$$u(z) = \left(\frac{z^2}{2} - hz\right)\frac{1}{\mu}\frac{\partial p}{\partial x}. \quad (12)$$

Conservation of mass requires that:

$$\frac{\partial h}{\partial t} + \frac{\partial}{\partial x}\int_0^h u(z)\, dz = 0. \quad (13)$$

The total discharge Q is obtained by integrating (12) across the film height:

$$Q = \int_0^h u(z)\, dz = -\frac{h^3}{3\mu}\frac{\partial p}{\partial x}, \quad (14)$$

which can then be inserted into (13) to give the nonlinear evolution equation for the fluid depth:

$$\frac{\partial h}{\partial t} = \frac{\partial}{\partial x}\left(\frac{h^3}{3\mu}\frac{\partial p}{\partial x}\right). \quad (15)$$

Integrating (5) with respect to z, subject to boundary condition (8), gives:

$$p = -\sigma \frac{\partial^2 h}{\partial x^2} + \rho g h. \quad (16)$$

**Non-Dimensionalization of the Lubrication Approximation Equations**

The governing equations for lubrication approximation for an incompressible Newtonian fluid are given by (15) and (16). Variables have been made dimensionless using the following scalings, where the tilde denotes dimensionless variables:

$$\tilde{h} = \frac{h}{H_0} \qquad \tilde{x} = \frac{x}{L_0} \qquad \tilde{t} = \frac{t}{T_0} \qquad \tilde{p} = \frac{p}{P_0}$$

The characteristic height $H_0$ represents the film thickness, and $L_0$ is the total length of the film across the substrate. The pressure and time scales are defined as:

$$P_0 = \frac{\sigma H_0}{L_0^2} \qquad T_0 = \frac{\mu L_0^2}{P_0 H_0^2} = \frac{\mu L_0^2}{H_0^2} \frac{L_0^2}{\sigma H_0} = \frac{\mu L_0}{\sigma \epsilon^3}$$

The non-dimensionalization of governing equations (15) and (16) leads to dimensionless equations (17) and (18) respectively.

$$3\frac{\partial \tilde{h}}{\partial \tilde{t}} = \frac{\partial}{\partial \tilde{x}}\left(\tilde{h}^3 \frac{\partial \tilde{p}}{\partial \tilde{x}}\right) \qquad (17)$$

$$\tilde{p} = -\frac{\partial^2 \tilde{h}}{\partial \tilde{x}^2} + Bo \cdot \tilde{h} \qquad (18)$$

The dimensionless Bond number is defined as the ratio of body force to surface tension effects:

$$Bo = \frac{\rho g L_0^2}{\sigma}.$$

The number should be enclosed in parentheses and set flush right in the column on the same line as the equation.

**Non-Dimensionalization of the Full 2D Navier-Stokes Equations**

The governing equations for a 2D dam break flow of an incompressible Newtonian fluid are given by (1), (2) and (3). Unlike in the lubrication approximation, variables have been made dimensionless using a single length and velocity scale, $L_0$ and $U_0$, for both spatial dimensions $x$ and $z$. Dimensionless variables are denoted with a tilde.

$$\tilde{u} = \frac{u}{U_0} \qquad \tilde{w} = \frac{w}{U_0} \qquad \tilde{x} = \frac{x}{L_0} \qquad \tilde{z} = \frac{z}{L_0} \qquad \tilde{t} = \frac{t}{T_0} = \frac{t}{L_0/U_0} \qquad \tilde{p} = \frac{p}{P_0}$$

We define the pressure scale $P_0 = \rho U_0^2$, Reynolds number $Re = \frac{\rho U_0 L_0}{\mu}$ and Froude number $Fr = \frac{U_0}{\sqrt{gL_0}}$ to obtain the dimensionless Navier-Stokes z-momentum equation from (2):

$$\left(\frac{\partial \tilde{w}}{\partial \tilde{t}} + \tilde{u}\frac{\partial \tilde{w}}{\partial \tilde{x}} + \tilde{w}\frac{\partial \tilde{w}}{\partial \tilde{z}}\right) = -\frac{\partial \tilde{p}}{\partial \tilde{z}} + \frac{1}{Re}\left(\frac{\partial^2 \tilde{w}}{\partial \tilde{x}^2} + \frac{\partial^2 \tilde{w}}{\partial \tilde{z}^2}\right) - \frac{1}{Fr^2}. \qquad (19)$$

Similarly, non-dimensionalisation of (1) gives the dimensionless Navier-Stokes x-momentum equation:

$$\left(\frac{\partial \tilde{u}}{\partial \tilde{t}} + \tilde{u}\frac{\partial \tilde{u}}{\partial \tilde{x}} + \tilde{w}\frac{\partial \tilde{u}}{\partial \tilde{z}}\right) = -\frac{\partial \tilde{p}}{\partial \tilde{x}} + \frac{1}{Re}\left(\frac{\partial^2 \tilde{u}}{\partial \tilde{x}^2} + \frac{\partial^2 \tilde{u}}{\partial \tilde{z}^2}\right). \quad (20)$$

Substituting dimensionless variables into (3) simply gives:

$$\frac{\partial \tilde{u}}{\partial \tilde{x}} + \frac{\partial \tilde{w}}{\partial \tilde{z}} = 0. \quad (21)$$

**Equivalence Between the Lubrication Approximation and Navier-Stokes Models**

In order to compare the lubrication approximation and Navier-Stokes models accurately, the dimensionless parameters between the two models must be related.

$$Re = \frac{\rho U_0 L_0}{\mu} \qquad Ca = \frac{\mu U_0}{\sigma} \qquad Fr = \frac{U_0}{\sqrt{gL_0}} \qquad Bo = \frac{\rho g L_0^2}{\sigma}$$

Using these definitions, it can be shown that:

$$Bo = \frac{Re \cdot Ca}{Fr^2}. \quad (22)$$

**NUMERICAL SIMULATIONS**

Both the lubrication approximation and Navier-Stokes models were solved using COMSOL Multiphysics 5.2 software, similar to the implementation used in [15]. A finite element technique was used to solve the equations. The lubrication approximation equations (17) and (18) were solved using the coefficient form PDE module of COMSOL Multiphysics, solving for the two dependent variables $h$ and $p$. The geometry was created as a horizontal interval from 0 to 1, divided into a total of 401 elements. A step function from $H_1$ to $H_g$ located along the x-axis at $L_1$ was used to define the initial condition for the free surface height.

The full Navier-Stokes equations (19), (20) and (21) were solved using the two-phase flow, moving mesh branch of COMSOL Multiphysics. The moving mesh technique has the advantage of keeping a sharp fluid interface at the free surface, compared with other two-phase flow methods such as level-set and phase field. This meant that precise results could be obtained for important variables including free surface height and velocity. The geometry of the dam-break problem was created using two adjacent rectangles of heights $H_1$ and $H_g$, and lengths $L_1$ and $L_g$ respectively, see Figure 1. The mesh used throughout was unstructured and consisted of approximately 7500 triangular elements.

**VALIDATION**

To verify that the Navier-Stokes and the lubrication approximation models were physically reasonable, the free surface velocity data from both models was compared with experimental data for silicone oil, in Figure 2. The velocity measurements were obtained using a process called Particle Tracking Velocimetry (PTV). PTV is a process in which individual particles are tracked within a fluid. Small beads are placed in the fluid to act as the particles to be tracked. Two-dimensional PTV was used for this experiment, in which a high intensity light sheet illuminates the beads in a single plane of the flow. High speed imaging at a constant frame rate is used to capture the displacement of each particle; this is then translated into a velocity measurement. Table 1 shows a summary of the geometric parameters and flow conditions used in the experiment. The dynamic viscosity for silicone oil was obtained using rheometer data, shown in Figures 3 and 4.

*Table 1. Flow conditions and geometry of experiment*

| $\rho$ | $\mu$ | $\sigma$ | $H_1$ | $H_g$ | $L_1$ | $L_g$ |
|---|---|---|---|---|---|---|
| $970 \; kg/m^3$ | $0.438 \; Pa \cdot s$ | $0.021 \; N/m$ | $0.009 m$ | $0.019 m$ | $0.636 m$ | $0.28 m$ |

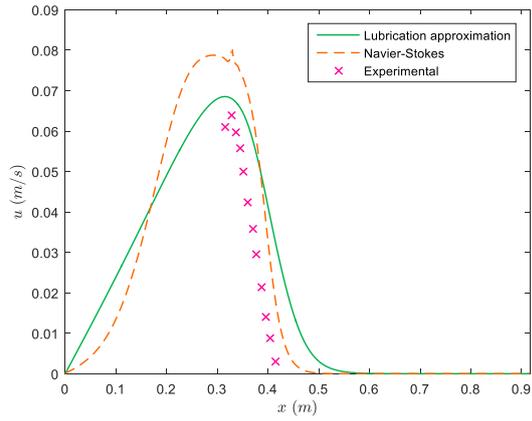
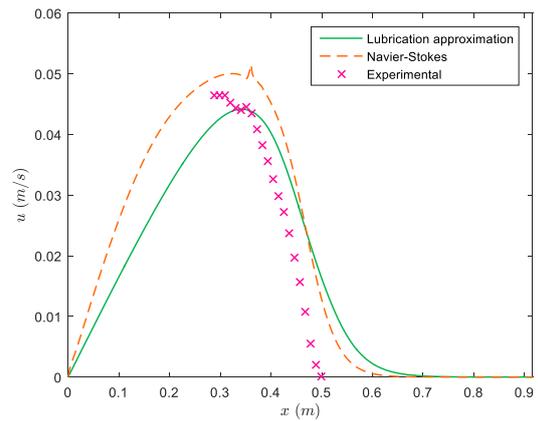

*(a)* *(b)*

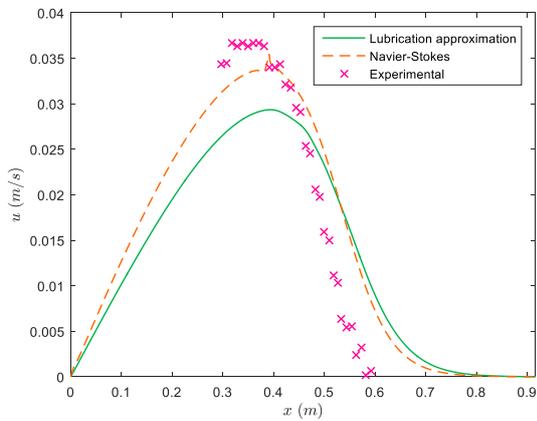
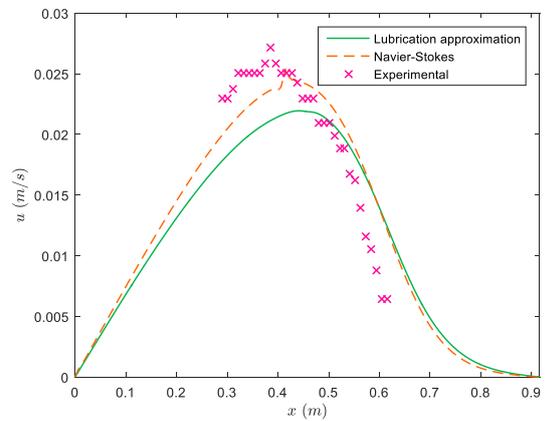

*(c)* *(d)*

*Figure 2. Comparison of experiment, lubrication approximation and Navier-Stokes models of free surface velocity at (a) t = 0.43s, (b) t = 0.93s, (c) t = 1.66s and (d) t = 2.46s*

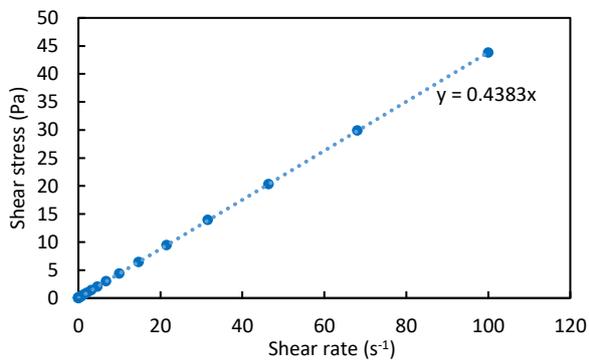
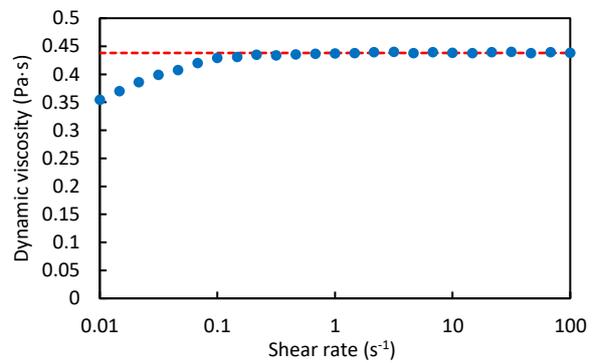

*Figure 3. Rheometer data for silicone oil*

*Figure 4. Rheometer data for silicone oil showing non-constant viscosity*

Figure 2 shows relatively good agreement between the experimental and theoretical data for free surface velocity. Discrepancies between the experiment and theory arose due to a range of inevitable differences. The gate was not removed instantaneously in the experiment as assumed in theory; it was moved upward through the fluid over a short period of time. The error decreases as this time approaches zero. The gate also had a thickness; it was not infinitely thin as the model supposes. The rheological parameters of the fluid used in the experiment had an associated uncertainty due to the shear rate-dependent viscosity, as demonstrated in Figure 4. The fluid behaves as a Newtonian for the most part; however, at low shear rates the viscosity changes by up to 20%.

The discrepancy between the two models observed in figure 2 was due to significant inertial effects produced by a relatively high aspect ratio of the film; this effect is analysed henceforth. In general, the results showed that the results obtained using the Navier-Stokes and lubrication approximation models were physically reasonable. The agreement provided the required confidence that the COMSOL modelling used was valid and could be used for further analyses.

**RESULTS AND DISCUSSION**

**Comparison of Lubrication Approximation and Navier-Stokes Solutions with Changing Re, Fr**

A parametric study of the dam-break problem was carried out for a fixed geometry and changing Reynolds and Froude numbers. The geometric parameters are recorded in table 5.1. Figure 5.1 shows a comparison of the free surface height for the lubrication approximation and Navier-Stokes models.

*Table 2. Geometric parameters*

| $H_1$ | $H_g$ | $L_1$ | $L_g$ |
|---|---|---|---|
| 0.0025 | 0.00575 | 0.7 | 0.3 |

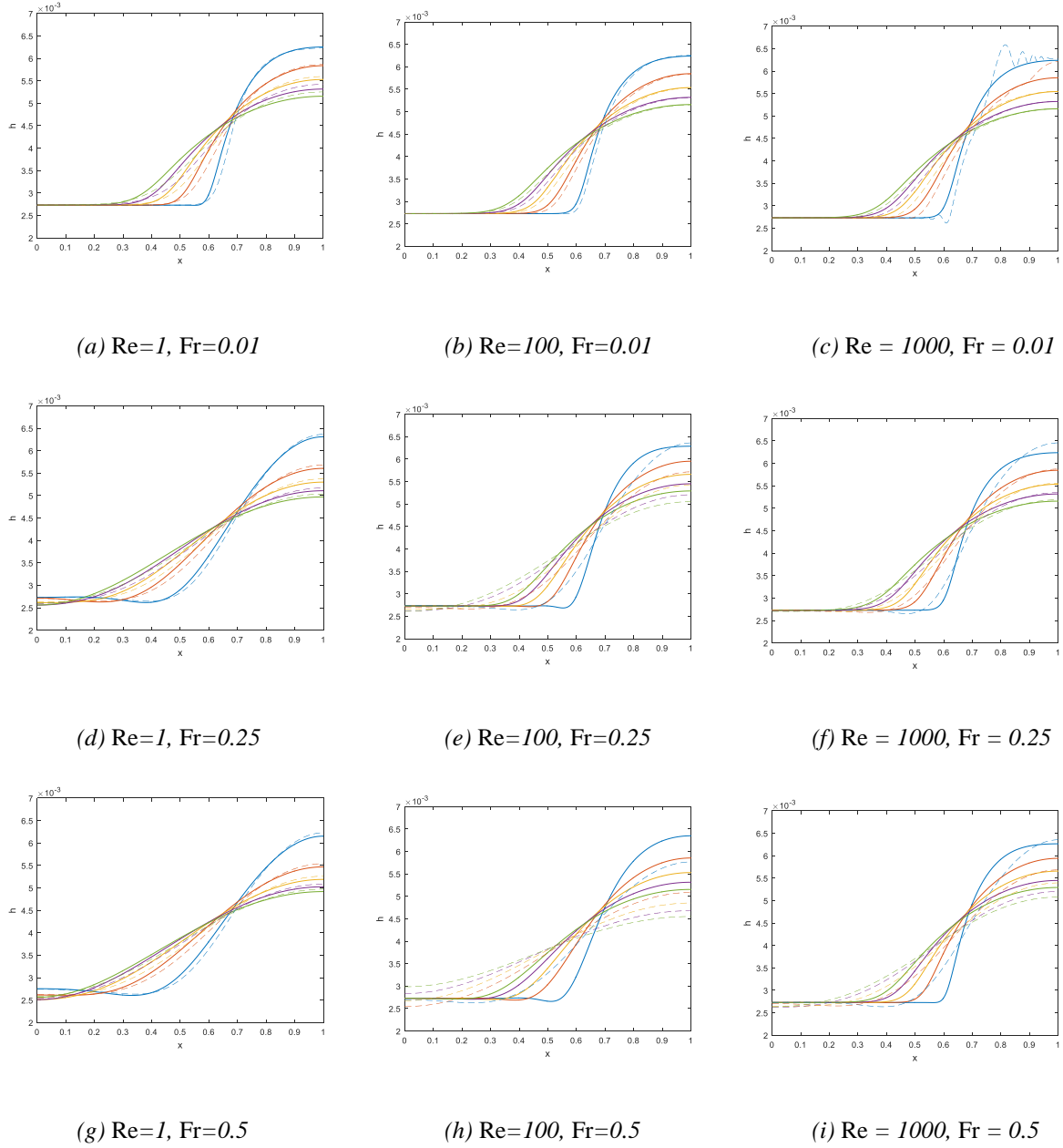

*(a) Re=1, Fr=0.01*  *(b) Re=100, Fr=0.01*  *(c) Re = 1000, Fr = 0.01*

*(d) Re=1, Fr=0.25*  *(e) Re=100, Fr=0.25*  *(f) Re = 1000, Fr = 0.25*

*(g) Re=1, Fr=0.5*  *(h) Re=100, Fr=0.5*  *(i) Re = 1000, Fr = 0.5*

*Figure 5. Dimensionless free surface height of film against dimensionless length across substrate at five dimensionless times with Re increasing horizontally across and Fr increasing vertically down the graphs. Dashed lines represent 2D Navier-Stokes (NS) solution and solid lines represent lubrication approximation (LA), with $x_{NS} = x_{LA}$, $h_{NS} = \epsilon h_{LA}$ and $\epsilon^3 t_{NS} = t_{LA}$ due to the different dimensionless scalings*

Figure 6 is a contour plot showing the constant error lines with changing dimensionless parameters Re and Fr. The percentage error $\delta$ is measured as the time-averaged mean absolute relative deviation (MARD) between the free surface height in the lubrication approximation and the Navier-Stokes models. ).

$$\delta = \frac{1}{t_f} \int_0^{t_f} \frac{1}{L_0} \int_0^{L_0} \left| \frac{h_{LA}(x,t) - h_{NS}(x,t)}{h_{NS}(x,t)} \right| dx\, dt \cdot 100\% \qquad (23)$$

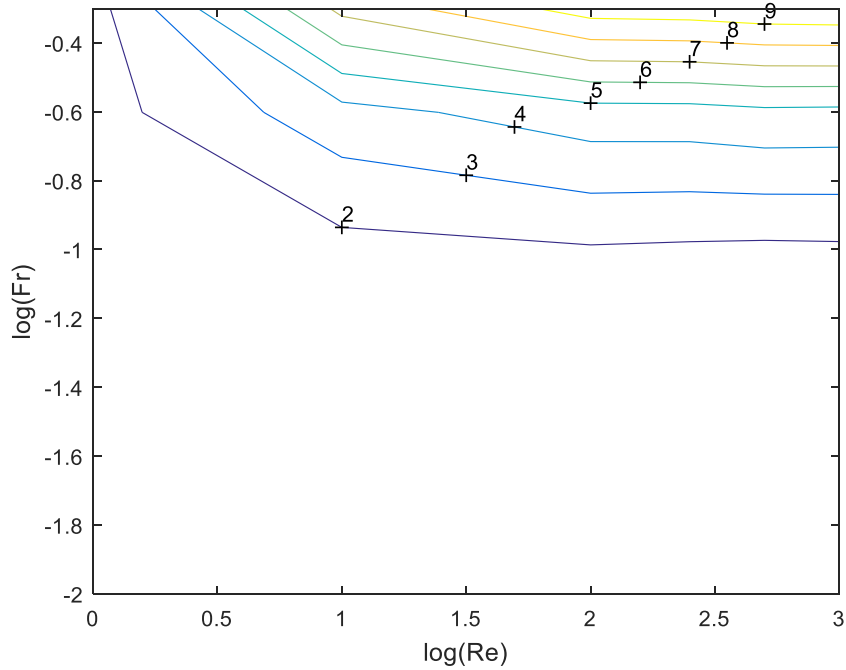

*Figure 6. Contours showing average percentage error of lubrication approximation compared with Navier-Stokes*

From Figures 5 and 6, it is evident that the error margin between the lubrication approximation and the Navier-Stokes solution increases with increasing Re and increasing Fr. The Reynolds number represents the ratio of fluid inertia to viscous forces, while the Froude number is defined as the ratio of inertia to gravity. Increasing both dimensionless groups has the effect of increasing the relative effect of fluid inertia in the levelling dynamics. As demonstrated in (4) and (5), inertial terms are neglected in the lubrication approximation. Hence, the result in Figure 6 is in agreement with the expected, that the lubrication approximation cannot capture the full levelling dynamics at high Re and Fr. This is true for all times during the levelling, and is evident particularly in Figure 5 (*h, i*) where the discrepancy between the two models is significant for all five dimensionless times shown. The discrepancies between the lubrication approximation and Navier-Stokes solutions at high Re and Fr can be visualised in Figures 7 and 8, showing the velocity fields at low and high Re and Fr respectively.

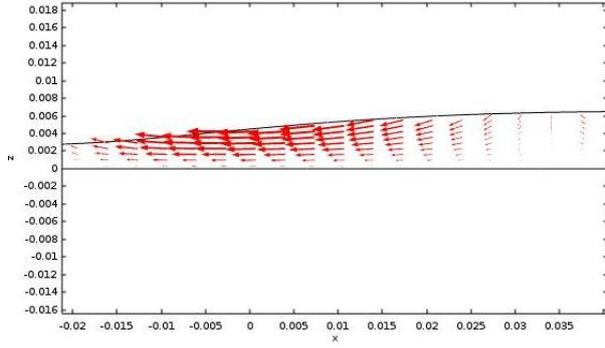 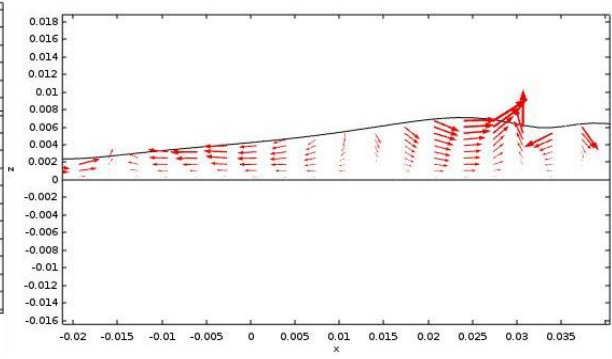

*Figure 7. Velocity field for* Re *= 100,* Fr *= 0.01*

*Figure 8. Velocity field for* Re *= 1000,* Fr *= 0.5*

At low Re and Fr, Figure 7 shows that the velocity profile is parabolic, which is in accordance with the model derived for the lubrication approximation (12). In contract, Figure 5.4 shows recirculation occurring during the levelling process, a flow feature which cannot be captured by the lubrication approximation. Hence, the discrepancy arises between the two models in Figure 5 (*i*). Interestingly, at extremely low Fr and high Re there appears to be less agreement between lubrication approximation and Navier-Stokes than at slightly higher Fr. Figure 5 (*c*) shows the formation of capillary waves at Re = 1000 and Fr = 0.01, observed in only the Navier-Stokes solution and not in the lubrication approximation model. The waves predominantly exist at early times in the levelling, smoothing out quickly as the film levels with time. At large times, unlike at high Fr and Re, the agreement between the lubrication approximation and Navier-Stokes model becomes acceptable. Hence, the overall error computed in (23) is not significant. However, if accurate results for the free surface height of the film are required for all times, the lubrication approximation would not be an acceptable model at early times in the levelling.

**Comparison of Lubrication Approximation and Navier-Stokes Solutions with changing aspect ratio $\epsilon$**

Figure 9 shows the effect of changing the film aspect ratio $\epsilon$ on the agreement between the lubrication approximation and Navier-Stokes solutions. The Reynolds and Froude numbers are fixed at 100 and 0.01 respectively. The ratio of $H_g/H_1$ was also kept constant at 2. The relationship between aspect ratio and the two film heights is demonstrated in (24). Figure 10 shows the percentage error of the lubrication approximation for each film aspect ratio.

$$\epsilon = \frac{H_0}{L_0} = \frac{H_1 + H_g}{2L_0} = \frac{3H_1}{2L_0} = \frac{3H_g}{4L_0} \tag{24}$$

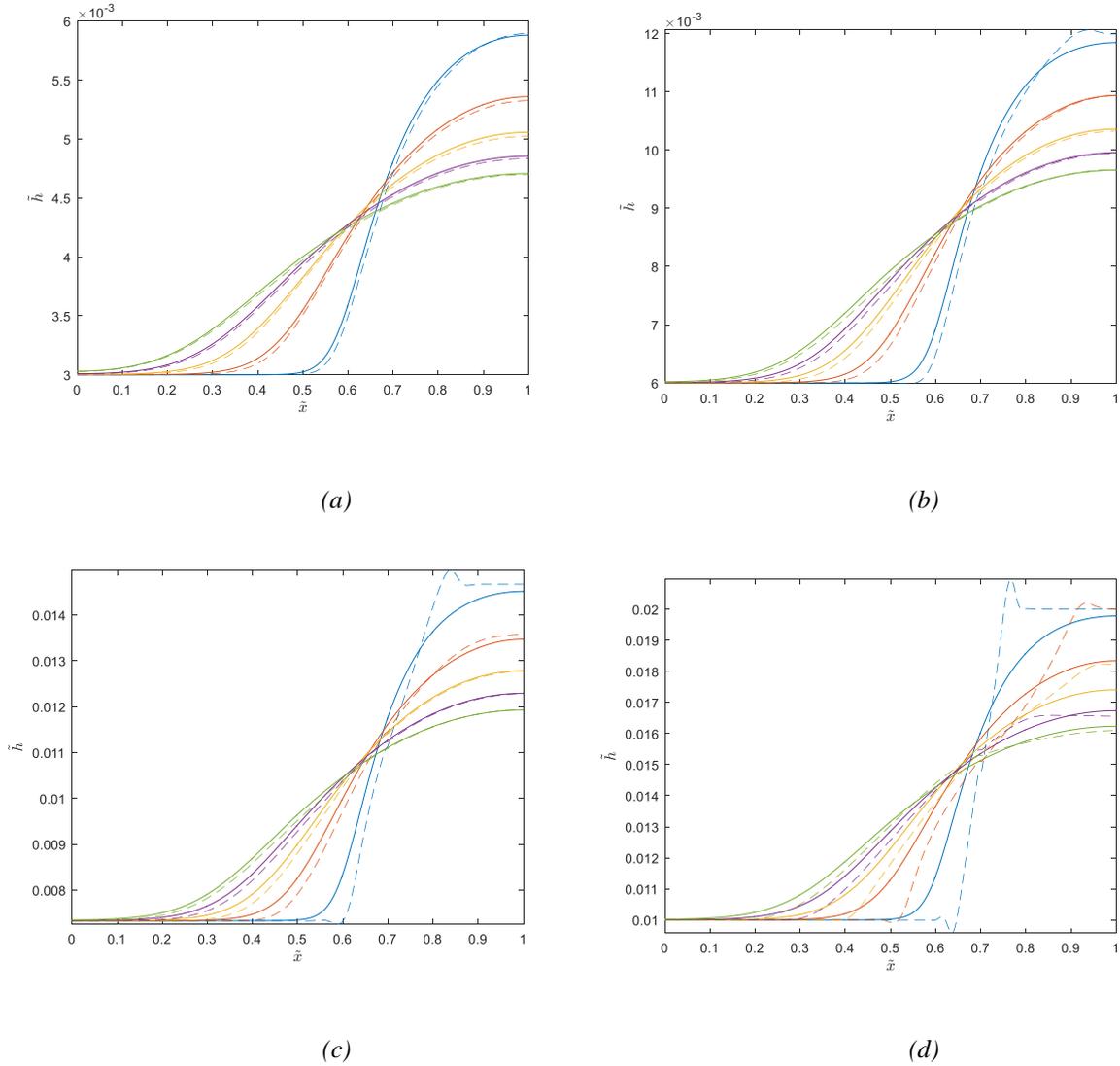

*Figure 9. Dimensionless free surface height of film against length across substrate at five dimensionless times with film aspect ratios (a) 0.0045, (b) 0.009, (c) 0.011, and (d) 0.015. Dashed lines represent Navier-Stokes solutions, solid lines represent lubrication approximation*

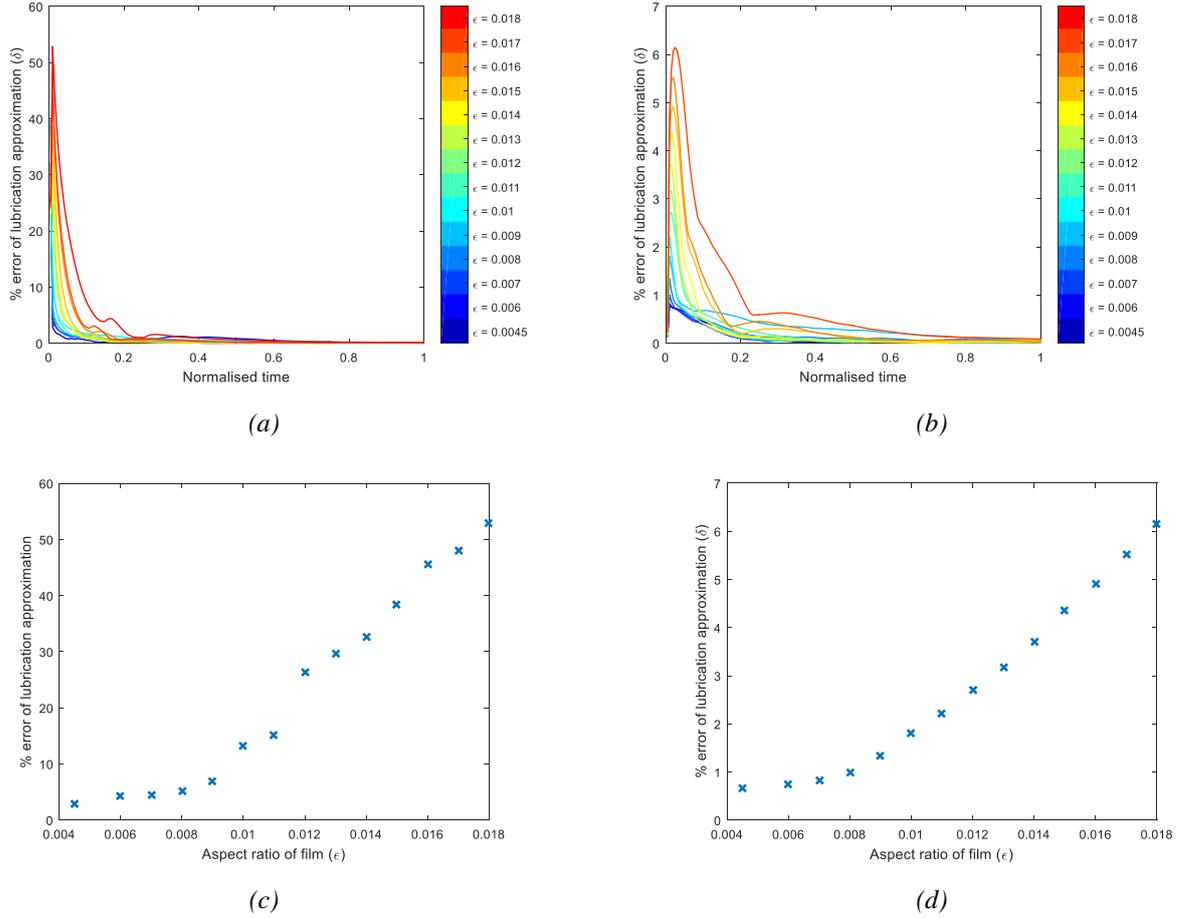

*Figure 10. Percentage error of free surface height using lubrication approximation compared to Navier-Stokes solution ($\delta$). The error is computed as the maximum percentage error in space for (a) and the average percentage error in space for (b). The maximum of the error over time in (a) and (b) is graphed against aspect ratio of the film ($\epsilon$) for (c) and (d) respectively. The normalised time in (a) and (b) is simply $t/t_{level}$, where $t_{level}$ is the total time taken for the fluid to level*

It can be seen from Figure 9 and Figure 10 that the error in the lubrication approximation decreases significantly with film aspect ratio, consistent with the lubrication theory assumption that $\epsilon \ll 1$. At high aspect ratios, specifically $\epsilon \geq 0.09$, capillary waves arise at the free surface and the levelling dynamics can no longer be accurately captured by the lubrication approximation. This result can be observed in Figure 9 (*c, d*) where the free surface height predicted by the lubrication approximation deviates significantly from the Navier-Stokes, particularly at early times in the levelling. At large times, similarly to the case with extremely low Fr in Figure 5 (*c*), the capillary waves level out and their effect on the free surface height becomes less significant. Therefore, the lubrication approximation is relatively accurate in predicting behaviour at long times even in cases where the film aspect ratio is large.

# CONCLUSIONS

In this work, the dam-break problem has been analysed for a thin film using two methods: the lubrication approximation, and a finite element solver for the 2D Navier-Stokes equations. The mathematical models for both methods consist of highly non-linear PDE's which are solved using a finite element technique. The models were validated using velocity data from dam-break experiments with silicone oil, which gave the required confidence that mathematical models and their numerical implementation were accurate. The results showed that in general, the lubrication approximation is effective in describing the dam-break problem for thin liquid films at low Reynolds and Froude numbers. High Re and Fr values implied higher fluid inertia, and the results showed that the lubrication approximation could not accurately capture the dynamics of the flow when inertial effects were significant. The full Navier-Stokes solution must be used to model the flow accurately at high Re and Fr values. Also at high film aspect ratios, the full Navier-Stokes solution must be used in order to capture capillary waves, which arise at approximately $\epsilon = 0.01$ for Reynolds and Froude numbers of 100 and 0.01 respectively.

# NOMENCLATURE

| | |
|---|---|
| $H_0, H_1, H_g$ | Average, lower, and upper liquid level, respectively |
| $L_0, L_1, L_g$ | Total span of the domain, span of lower level liquid layer, and span of the upper level liquid layer, respectively |
| $Re, Fr, Bo, Ca$ | Reynolds number, Froude number, Bond number, Capillary number, respectively. |
| $\varepsilon$ | Film aspect ratio |
| $\mu, \rho, \sigma, g$ | Dynamic viscosity, density, surface tension, and acceleration of gravity, respectively |
| $(u, v, w)$ | Velocity vector |
| $(x, y, z), t$ | Position vector and time, respectively |
| $h(x,t), h_{LA}(x,t), h_{NS}(x,t)$ | Film thickness, film thickness from lubrication |

| | approximation, and film thickness from Navier-Stokes, respectively |
|---|---|
| $Q(x,t)$ | Flow rate |
| $\delta$ | Percentage relative error |